\documentclass[twoside]{article}

\usepackage{amssymb,latexsym,amsmath}     
\usepackage[utf8]{inputenc} 
\usepackage[T1]{fontenc}    
\usepackage{hyperref}       
\usepackage{url}            
\usepackage{booktabs}       
\usepackage{amsfonts}       
\usepackage{nicefrac}       
\usepackage{microtype}      
\usepackage{lipsum}
\usepackage{fancyhdr}       
\usepackage{graphicx}       
\usepackage{array}          
\usepackage{tabularx}       
\graphicspath{{media/}}     

\title{Quantitative 3D Analysis of Porosity and Fractal Geometry in Electrochemically Etched Macroporous Silicon}

\author{
  A. Ramírez-Porras$^{1}$\thanks{Corresponding author: \texttt{arturo.ramirez@ucr.ac.cr}},
  I. Prado$^{1}$,
  N. R. Schwarz$^{2}$,
  U. Steiner$^{2}$\\[1ex]
  \small $^{1}$CICIMA and Escuela de Física, Universidad de Costa Rica\\
  \small $^{2}$Adolphe Merkle Institute, University of Fribourg, Switzerland
}

\begin{document}
\maketitle

\begin{abstract}
Macroporous silicon is widely employed in sensing and optoelectronic applications due to its large internal surface area and adjustable pore structure. However, quantitative correlations between morphology and functionality require accurately characterizing the three-dimensional pore network. In this study, we used focused Ga$^{+}$ ion beam--scanning electron microscopy (FIB--SEM) tomography to reconstruct representative volumes of electrochemically etched macroporous silicon layers. We extracted true three-dimensional porosity and surface-to-volume ratios and compared them with two-dimensional estimates obtained from SEM images. Our results demonstrate that surface-based porosity systematically underestimates true volumetric porosity. These discrepancies arise from anisotropy, branching, and variability in pore-size. Fractal analysis reveals that the pore network has moderate geometric complexity, consistent with electrochemical macropore formation mechanisms. The results highlight the importance of direct 3D characterization for reliable morphological quantification and provide a robust framework for interpreting structural trends in macroporous silicon.
\end{abstract}

\section{Introduction}
Porous silicon (PSi) has emerged as a scientifically and technologically significant material since its efficient photoluminescence was discovered in the 1990s \cite{Canham1990}. PSi’s combination of a high surface area, tunable pore morphology, and compatibility with standard silicon processing makes it as a promising candidate for a wide range of applications. PSi has been extensively studied for use in advanced optoelectronic devices, such as light-emitting diodes and photovoltaic cells, where its nanostructure can manipulate light \cite{Bisi2000}. Additionally, the large, accessible internal surface of PSi makes it an exceptional platform for chemical and biological sensing. When target analytes are adsorbed or infiltrated into the porous matrix, the optical and electrical properties of the material (e.g., refractive index) change, providing a highly sensitive transduction mechanism \cite{Sailor2012}. Macroporous silicon, characterized by pore diameters on the order of microns, is particularly relevant for sensor development. Its larger pore structures facilitate the infiltration of larger biomolecules (e.g., proteins) and enable faster response times. These properties make microporous silicon suitable for biosensing and microfluidic integration \cite{Lin1997}.

The performance of PSi in these applications is intrinsically governed by its nanoscale and microscale morphology. Key parameters such as total internal surface area, pore size distribution, tortuosity, and, critically, the surface-to-volume ratio (S/V) within the porous layer directly determine its optical, electrical, and catalytic properties. For sensors, the S/V ratio correlates directly with the density of available binding sites and the device's ultimate sensitivity. Historically, morphology assessment has relied on two-dimensional analysis of cross-sectional or planar scanning electron microscopy (SEM) images. Software like ImageJ or Fiji is commonly used to estimate 2D porosity (pore area fraction) and, through stereological assumptions, infer surface density from these images \cite{Schneider2012}. While these two-dimensional methodologies are useful for comparative studies, they are inherently limited. They assume structural homogeneity and isotropic geometry, which is often an oversimplification for complex porous networks. These assumptions lead to significant errors when estimating true volumetric parameters, such as the surface-to-volume (S/V) ratio \cite{Rossberg2025, Maire2014}.

Conventional methods for measuring porosity, such as gravimetric analysis, only provide an average bulk value. These methods do not provide information about the spatial distribution of pores or the connectivity of the porous network \cite{Halimaoui1997}. Reliably linking morphology to function requires true three-dimensional morphological characterization. Focused ion beam scanning electron microscopy (FIB-SEM) tomography is a key technique for nanoscale 3D analysis. In this method, a focused gallium ion beam sequentially mills away thin slices of material (typically 10–50 nm thick), simultaneously imaging freshly exposed cross-sections with an electron beam after each milling step \cite{Giannuzzi2005}. This serial sectioning generates a stack of aligned 2D images that can be reconstructed into a precise digital 3D volume of the material.

FIB--SEM tomography offers three key advantages for characterizing porous silicon: First, it provides direct, unambiguous measurements of the 3D pore network. These measurements enable accurate calculations of the S/V ratio and tortuosity, eliminating the need for stereological assumptions. Second, FIB-SEM achieves high resolution, typically below 10 nm in both lateral dimensions and in slice thickness. This capability can resolve fine features in mesoporous and macroporous layers. Third, FIB-SEM tomography enables the visualization of structural anisotropy and defects that are invisible to indirect methods \cite{Inkson2016}. This study uses Ga$^{+}$ FIB--SEM tomography to conduct a quantitative, 3D morphological analysis of macroporous silicon layers intended for sensor applications. We precisely extract of volumetric parameters, including the true S/V ratio, and contrast these findings with values derived from conventional 2D image analysis. These results highlight the importance of direct 3D characterization in the rational design and optimization of porous silicon-based devices.

Accurate three-dimensional morphology is also essential for predicting the functional behavior of porous silicon in sensing and optoelectronic applications. Device responses such as optical interference, refractive-index modulation upon infiltration, and analyte transport depend strongly on pore connectivity, tortuosity, and internal surface distribution \cite{Bisi2000, Sailor2012, Lin1997}. These factors cannot be reliably inferred from 2D projections, which obscure branching and anisotropy and therefore lead to misestimation of transport paths and effective refractive indices. By resolving the full pore network, 3D reconstructions enable more reliable predictions of adsorption kinetics, optical contrast, and infiltration behavior, directly improving the interpretation of measurement results and performance modeling in PSi-based sensors and photonic structures \cite{Sailor2012, Maire2014}.

\section{Experimental Details}
A total of four PSi samples were fabricated using a top-down electrochemical etching approach. The starting material was a boron-doped, p-type crystalline silicon wafer that was 10 cm diameter and 300 µm thick. The wafer was diced into square specimens measuring approximately 1 cm$^{2}$. Each specimen was affixed to the base of a custom Teflon electrochemical cell containing a circular aperture with an exposed area of 0.13 cm$^{2}$  that defined the active etching region. Electrochemical etching was carried out using an aqueous hydrofluoric acid (HF) solution in an ethanol, prepared at two distinct HF concentrations: 12.5 \% and 20 \% (v/v). A platinum (Pt) rod served as the counter electrode (cathode) and was immersed directly in the etchant. Electrical contact to the silicon substrate (anode) was established by pressing a piece of aluminum foil against the rear surface of the silicon sample. A constant-current power supply applied one of two current densities—corresponding to 23.1 mA/cm$^2$ or 53.8 mA/cm$^2$—across the electrochemical cell to initiate pore formation. The etching duration was set to 20, 30, and 40 minutes for different samples. After each etching step, the samples were removed from the cell and rinsed sequentially in ethanol and deionized water to remove residual etchant and reaction byproducts. After preparing a total of 20 samples with different parameters, the suitable PSi samples for the study were selected based on the combination of HF concentration, applied current and etching time: (S1) 12.5 \% HF at 23.1 mA/cm$^2$ for 40 min; (S2) 20 \% HF at 23.1 mA/cm$^2$ for 40 min; (S3) 12.5 \% HF at 53.8 mA/cm$^2$ for 30 min; (S4) 20 \% HF at 53.8 mA/cm$^2$ for 20 min; and (S5) 20 \% HF at 53.8 mA/cm$^2$ for 40 min.

We obtained images of the fabricated PSi samples using a scanning electron microscopy (SEM) (Jeol, model JSM IT1500). Before inserting the samples into the FIB-SEM system, we treated them with a low-viscosity, UV curable resin (product No.900149, Aldrich) and placed in a spin coater system (model WS-650MZ-23NPPB, Laurel Tehnologies Corporation), ramping up to 4000 rpm over 30 s. Then, we stored samples in a desiccator for at least four hours and irradiated them with 254 nm UV light at approximately 126 W/cm$^{2}$ for 180 seconds.

We next sputtered a 4 nm gold coating onto each sample using a Cressington 208HR sputter coater to ensure good conductivity for the SEM. We attached the specimens to a conductive sample holder and placed them inside the vacuum chamber of a FIB-SEM (ThermoScientific, model Scios 2 Dual Beam). After selecting a suitable region on the sample surface, we determined the eucentric condition at a working distance of 7 mm and a sample tilt of 52°. Using a gas injection system (GIS), we deposited a 500 nm-thick platinum protective layer with horizontal dimensions of 20 or 30 µm long and 10 to 20 µm wide on top of the Au-coated resin layer. A coarse trench was milled in front of the selected region to expose the imaging surface, and side trenches were added to collect sputtered material and reduce redeposition. A fiducial marker was created to minimize drift from beam and stage movements. Serial sectioning was automated using ThermoFisher Auto Slice and View (v. 4). Slices of 250 nm thick were milled with a Ga$^{+}$ beam at 30 kV and 0.30 nA. After each slice, images were acquired in OptiTilt mode using the Everhart–Thornley (for secondary electrons) and in-lens T1 (for back-scattered electrons) detectors. Tilt-related distortion from the 52° acquisition angle was corrected using the system’s built-in tilt compensation. The image stacks were processed using the T1 image sets, which provided superior contrast, in ilastik (www.ilastik.org) to generate well-segmented silicon–resin image sequences. Three-dimensional reconstructions were produced using Fiji’s built-in 3D Viewer plugin. Fractal dimensions and pore-radius ratios were quantified with the BoneJ plugin in Fiji.

\section{Results and Discussion}
Table 1 summarizes the thickness values obtained from the analysis of cross-sectional images of samples S1–S5. The reported thicknesses correspond to mean values extracted from multiple measurements along the sample cross sections to ensure statistical robustness.

\begin{table}[h]
\centering
\caption{Mean thickness (T) measurements of the different samples according to the synthesis parameters: current density and HF concentration.}
\begin{tabular}{ | m{2.25cm} | m{4.25cm}| m{4.0cm} | }
\toprule
Curr. Density (mA/cm$^2$) & 12.5\% HF conc. & 20.0\% HF conc. \\
\midrule
23.1 & S1 (40 min): T=29.8 $\mu m$ & S2 (40 min): T=51.2 $\mu m$\\
\midrule
53.8 & S3 (30 min): T=64.1 $\mu m$ & S4 (20 min): T=51.2 $\mu m$\
S5 (40 min): T=105.8 $\mu m$\\
\bottomrule
\end{tabular}
\end{table}

Figure 1 shows the top view and cross-section of sample S3 as an example. As expected, the thickness increases with etching time and current density, although there are some deviations from monotonicity. This behavior suggests that local mass transport and electrolyte depletion effects may also play a significant role in electrochemical etching.

\begin{figure} 
    \centering
    \includegraphics[width=1.0\linewidth]{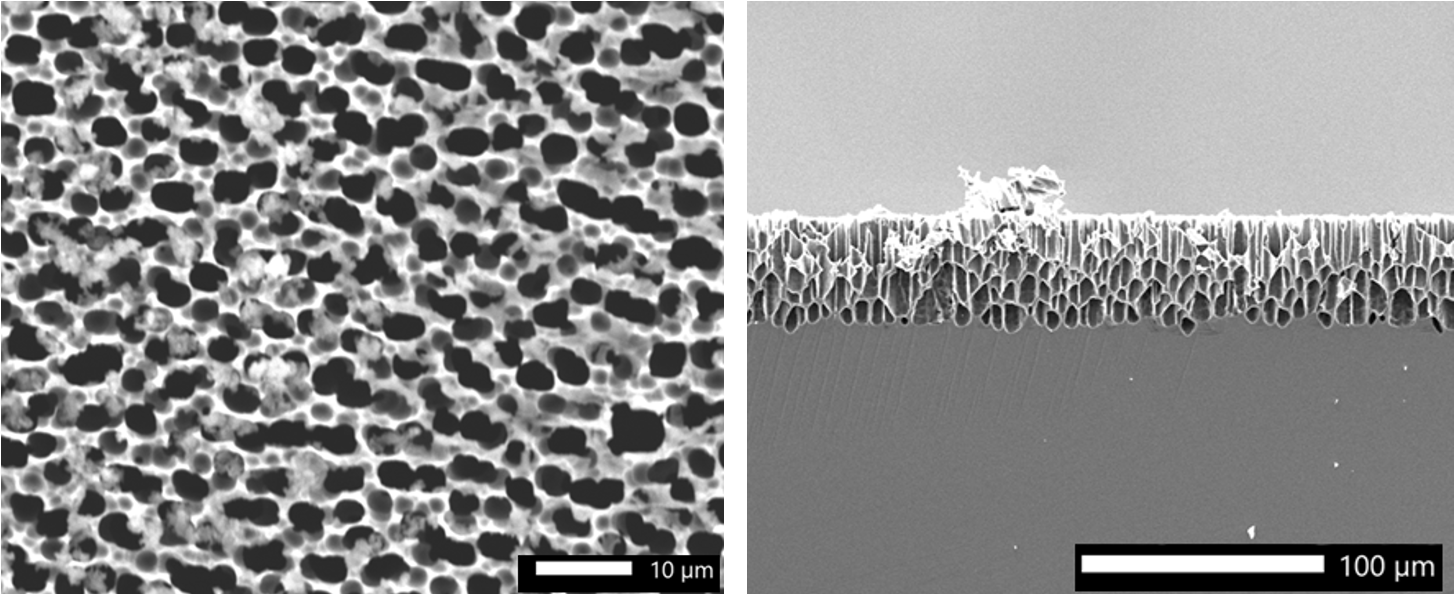}
    \caption{Top view (left) and cross section (right) of the S3 sample.}
    \label{fig:Fig1}
\end{figure}

Three-dimensional tomography renderings of representative samples (Fig. 2) reveal a highly interconnected, macroporous network extending through the silicon layer. Image segmentation and reconstruction were performed using established, open-source tools to ensure reproducibility. The rendered volumes confirm that the pore morphology is continuous and anisotropic, oriented preferentially along the etching direction. These findings are consistent with previous reports on electrochemically etched macroporous silicon.

\begin{figure} 
    \centering
    \includegraphics[width=1.0\linewidth]{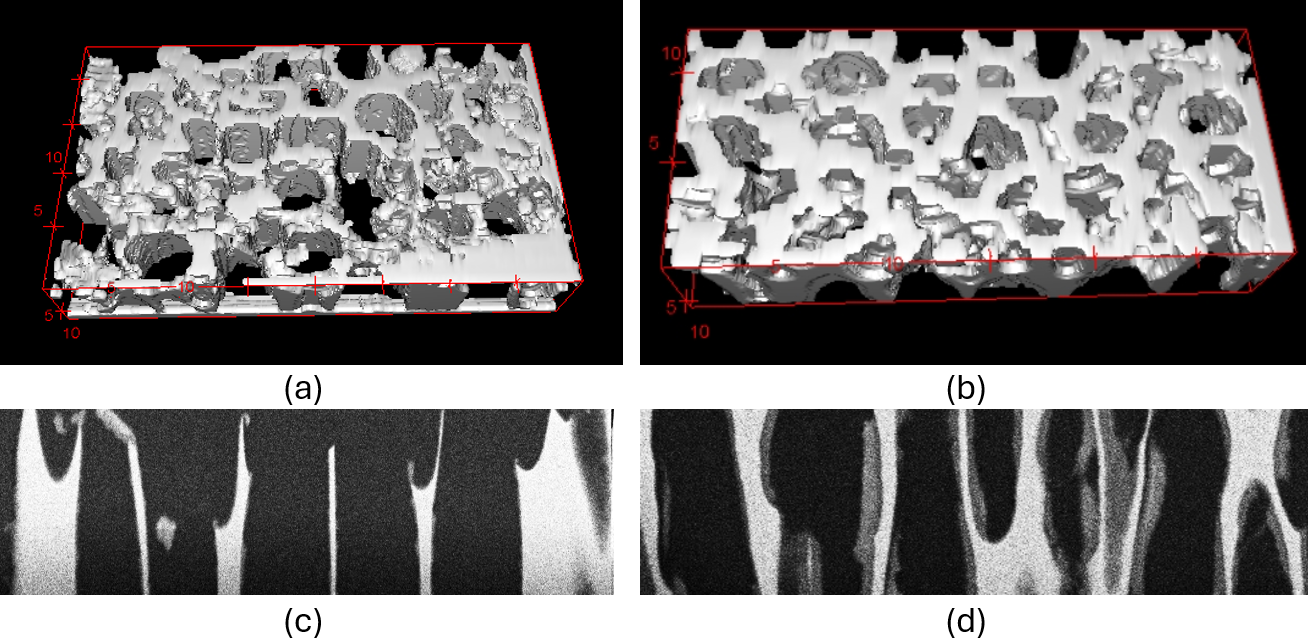}
    \caption{3D tomography renderings of S1 (a) and S4 (b). The dimensions of the renderings are 40 $\mu m$ (length), 24.5 $\mu m$ (width), 7 $\mu m$ (thickness) for S1, and 30 $\mu m$ (length), 15 $\mu m$ (width), 6.1 $\mu m$ (thickness) for S4. (c) and (d) are actual tomography images taken near the center of samples S1 and S4, respectively, with the SEM during the tomography run. The pores are shown in black, and the Si material is shown in white.}
    \label{fig:Fig2}
\end{figure}

The three-dimensional porosity was quantified by the ratio of the silicon skeleton’s volume to the total volume, according to:

\begin{equation}
   \phi = 1 - \frac{V_{\text{Si}}}{V_{\text{total}}}. 
\end{equation}

Table 2 compares the two-dimensional porosity values ($\phi_{2D}$) calculated using Fiji on the surface of the region analyzed for three-dimensional porosity ($\phi_{3D}$) for the studied samples, and the ratio between the two. In all cases, $\phi_{2D}<\phi_{3D}$, indicating that 2D surface analysis systematically underestimates the true porosity of the bulk pore network. This discrepancy reflects the anisotropic and interconnected nature of the pores and demonstrates that surface views alone are insufficient to capture the full three-dimensional morphology.

\begin{table}[h]
\centering
\caption{Comparison of 2D and 3D porosity values of the studied samples. The fractal dimension is also included, as well as parameter $\rho$ defined in Eq. [2]. The current densities, HF concentrations and etching times are displayed in parentheses.}
\begin{tabular}{ | m{6.0cm} | m{1.0cm}| m{1.0cm} | m{1.25cm} | m{1.0cm} | m{1.0cm} | }
\toprule
Sample (Curr. dens.; [HF]; Etch. time) & $\phi_{2D}$ & $\phi_{3D}$ & $\phi_{2D}/\phi_{3D}$ & $D_f$ & $\rho$ \\
\midrule
S1 (23.1 mA/cm$^2$; 12.5\%; 40 min) & 0.62 & 0.72 & 0.87 & 2.51 & 0.51 \\
S2 (23.1 mA/cm$^2$; 20.0\%; 40 min) & 0.49 & 0.50 & 0.98 & 2.50 & 0.27 \\
S3 (53.8 mA/cm$^2$; 12.5\%; 30 min) & 0.52 & 0.66 & 0.78 & 2.37 & 0.52 \\
S4 (53.8 mA/cm$^2$; 20.0\%; 20 min) & 0.37 & 0.61 & 0.60 & 2.47 & 0.39 \\
S5 (53.8 mA/cm$^2$; 20.0\%; 40 min) & 0.47 & 0.51 & 0.91 & 2.61 & 0.16 \\
\bottomrule
\end{tabular}
\end{table}

If the pore system consisted of non-interacting, isotropically distributed cylindrical pores, then $\phi_{2D}$ and $\phi_{3D}$ would coincide. However, the observed ratios range from 0.60 to 0.91, indicating asymmetry, branching, and variations in pore diameter within the analyzed volumes. The extracted fractal dimensions ($D_f$), ranging from approximately 2.37 to 2.61, indicate that the pore network exhibits scale-dependent geometric complexity. These values are consistent with those of macroporous silicon formed under electrochemical conditions and with the results of previous fractal analyses of porous silicon structures. Within a fractal pore-space framework, porosity is related to the ratio of the minimum to maximum characteristic pore radii as follows \cite{Yu2002,Sundararajan2022}:

\begin{equation}
\phi_{3D} = \left(\frac{r_{\min}}{r_{\max}}\right)^{3-D_f} = \rho^{3-D_f}
\end{equation}

This relation was used to calculate the parameter $\rho=r_{min}/r_{max}$ for all samples. The resulting r values range from 0.16 to 0.52, indicating that the minimum pore radius is typically only slightly smaller than the maximum radius. This relatively narrow distribution confirms that the macroporous network does not exhibit extreme multiscale disparity despite the presence of heterogeneity and branching. High-contrast SEM images (Fig. 3) qualitatively support this conclusion by revealing coexisting pores of different sizes within the same field of view.

\begin{figure} 
    \centering
    \includegraphics[width=1.0\linewidth]{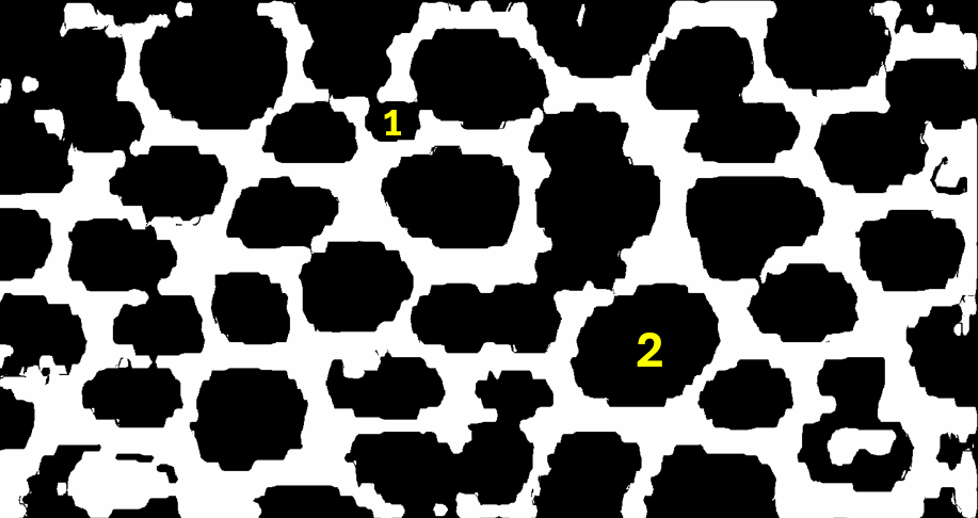}
    \caption{High-contrast image of sample S1. The image has been obtained by taking the first few layers of the 3D render and enhancing the contrast using the thresholding function of Fiji. The horizontal width is 40 $\mu m$. The pores are shown in black, and the Si material is shown in white. The mean radius ratio between the smaller pore (marked 1) and the larger pore (marked 2) is 0.34, which falls within the range of 0.16 to 0.52.}
    \label{fig:Fig3}
\end{figure}

Overall, combined porosity and fractal analysis demonstrate that the morphology of electrochemically etched macroporous silicon is characterized by moderate geometric complexity rather than extreme disorder.

\section{Conclusions}
We used Ga$^{+}$ FIB--SEM tomography to characterize the three-dimensional structure of electrochemically etched, macroporous silicon. The reconstructed volumes reveal a highly interconnected, anisotropic pore network that cannot be accurately described by conventional two-dimensional image. Quantitative comparison between 2D and 3D porosity values shows that surface-based methods systematically underestimate true porosity. These deviations reflect pore branching and diameter variability.

Fractal analysis indicates that the layers of macroporous silicon exhibit moderate geometric complexity, with fractal dimensions ranging from approximately 2.4 to 2.6. These values are consistent with electrochemical pore growth models and confirm that the pore system is neither perfectly ordered nor strongly disordered. Extracted pore radius ratios demonstrate a relatively narrow distribution of characteristic length scales despite morphological heterogeneity.

Overall, this study establishes FIB–SEM tomography as a powerful and reliable tool for quantitatively analyzing the morphology of macroporous silicon. The methodology and results presented here provide a solid basis for future investigations linking three-dimensional morphology to transport, optical, and sensing properties in porous silicon-based devices.

\section*{Acknowledgments}
This work was partially supported by the Vicerrectoría de Investigación of the Universidad de Costa Rica. We also acknowledge funding from the Swiss National Science Foundation (grant number IZSEZ0-237437) and from the Adolphe Merkle Foundation.

\bibliographystyle{unsrt}
\bibliography{References}

@article{Canham1990,
  author    = {L. T. Canham},
  title     = {Silicon quantum wire array fabrication by electrochemical and chemical dissolution of wafers},
  journal   = {Applied Physics Letters},
  volume    = {57},
  number    = {10},
  pages     = {1046--1048},
  year      = {1990},
  doi       = {10.1063/1.103561}
}

@article{Bisi2000,
  author    = {O. Bisi and S. Ossicini and L. Pavesi},
  title     = {Porous silicon: a quantum sponge structure for silicon based optoelectronics},
  journal   = {Surface Science Reports},
  volume    = {38},
  number    = {1--3},
  pages     = {1--126},
  year      = {2000},
  doi       = {10.1016/S0167-5729(99)00012-6}
}

@book{Sailor2012,
  author    = {M. J. Sailor},
  title     = {Porous Silicon in Practice: Preparation, Characterization and Applications},
  publisher = {Wiley-VCH},
  address   = {Weinheim},
  year      = {2012},
  doi       = {10.1002/9783527641901}
}

@article{Lin1997,
  author    = {V. S.-Y. Lin and K. Motesharei and K.-P. S. Dancil and M. J. Sailor and M. R. Ghadiri},
  title     = {A Porous Silicon-Based Optical Interferometric Biosensor},
  journal   = {Science},
  volume    = {278},
  number    = {5339},
  pages     = {840--843},
  year      = {1997},
  doi       = {10.1126/science.278.5339.840}
}

@article{Schneider2012,
  author    = {C. A. Schneider and W. S. Rasband and K. W. Eliceiri},
  title     = {NIH Image to ImageJ: 25 years of image analysis},
  journal   = {Nature Methods},
  volume    = {9},
  number    = {7},
  pages     = {671--675},
  year      = {2012},
  doi       = {10.1038/nmeth.2089}
}

@article{Rossberg2025,
  author    = {N. Rossberg and S. Corrie and L. Gr{\o}ndahl and I. Jayawardena},
  title     = {Automated analysis of pore structures in biomaterials},
  journal   = {Journal of Materials Chemistry B},
  volume    = {13},
  pages     = {9377},
  year      = {2025},
  doi       = {10.1039/d5tb00848d}
}

@article{Maire2014,
  author    = {E. Maire and P. J. Withers},
  title     = {Quantitative X-ray tomography},
  journal   = {International Materials Reviews},
  volume    = {59},
  number    = {1},
  pages     = {1--43},
  year      = {2014},
  doi       = {10.1179/1743280413Y.0000000023}
}

@incollection{Halimaoui1997,
  author    = {A. Halimaoui},
  title     = {Porous silicon formation by anodisation},
  booktitle = {Properties of Porous Silicon},
  editor    = {L. Canham},
  publisher = {INSPEC},
  address   = {London},
  pages     = {12--22},
  year      = {1997},
  isbn      = {0-85296-932-5}
}

@book{Giannuzzi2005,
  author    = {L. A. Giannuzzi and F. A. Stevie},
  title     = {Introduction to Focused Ion Beams: Instrumentation, Theory, Techniques and Practice},
  publisher = {Springer US},
  address   = {Boston, MA},
  year      = {2005},
  doi       = {10.1007/b101190}
}

@incollection{Inkson2016,
  author    = {B. J. Inkson},
  title     = {Scanning electron microscopy (SEM) and transmission electron microscopy (TEM) for materials characterization},
  booktitle = {Materials Characterization Using Nondestructive Evaluation (NDE) Methods},
  editor    = {G. H{\"u}bschen and I. Altpeter and R. Tschuncky and H.-G. Herrmann},
  publisher = {Woodhead Publishing},
  pages     = {17--43},
  year      = {2016},
  doi       = {10.1016/B978-0-08-100040-3.00002-X}
}

@article{Yu2002,
  author    = {B. Yu and P. Cheng},
  title     = {A fractal permeability model for bi-dispersed porous media},
  journal   = {International Journal of Heat and Mass Transfer},
  volume    = {45},
  pages     = {2983},
  year      = {2002},
  doi       = {10.1016/S0017-9310(02)00014-5}
}

@incollection{Sundararajan2022,
  author    = {T. Sundararajan},
  title     = {Permeability in Pores},
  booktitle = {Transport Phenomena in Porous Media},
  publisher = {Springer},
  year      = {2022},
  doi       = {10.1007/978-3-030-83455-3_9}
}

\end{document}